\documentclass[prc,superscriptaddress,noshowpacs,unsortedaddress,twocolumn,showpacs,preprintnumbers,amsmath,amssymb]{revtex4-1}

\usepackage[dvipdfmx]{graphicx}
\usepackage{amsmath,amssymb,times}
\usepackage{color}
\usepackage{ulem}
\usepackage{bm}
\usepackage{here}

\newcommand{\beq}{\begin{equation}}
\newcommand{\eeq}{\end{equation}}
\newcommand{\bea}{\begin{eqnarray}}
\newcommand{\eea}{\end{eqnarray}}

\begin{document}
\title{Determination of matter radius and neutron skin of $^{58}$Ni \\
from  reaction cross section of proton+$^{58}$Ni scattering \\
based on chiral $g$-matrix model}

\author{Shingo~Tagami}
\affiliation{Department of Physics, Kyushu University, Fukuoka 819-0395, Japan}

\author{Maya~Takechi}
\affiliation{Niigata University, Niigata 950-2181, Japan}

\author{Jun~Matsui}
\affiliation{Department of Physics, Kyushu University, Fukuoka 819-0395, Japan}

\author{Tomotsugu~Wakasa}
\affiliation{Department of Physics, Kyushu University, Fukuoka 819-0395, Japan}

\author{Masanobu Yahiro}
\email[]{orion093g@gmail.com}
\affiliation{Department of Physics, Kyushu University, Fukuoka 819-0395, Japan}             

\date{\today}

\begin{abstract}
\noindent 
{\bf Background:}
Using the chiral  (Kyushu) $g$-matrix folding model with 
the densities calculated with Gogny-HFB (GHFB) with the angular momentum  projection (AMP), 
we determined the central values of matter radius  and neutron skin from the central values of 
reaction cross sections $\sigma_{\rm R}({\rm EXP})$ 
of p+$^{40,48}$Ca  and  p+$^{208}$Pb scattering. 
As for p+$^{58}$Ni scattering, $\sigma_{\rm R}({\rm EXP})$ are available 
as a function of incident energy $E_{\rm in}$.   
 \\
{\bf Aim:} 
Our aim is to determine  matter radius $r_{m}$ and skin $r_{\rm skin}$ for $^{58}$Ni  from the 
$\sigma_{\rm R}({\rm EXP})$ of  p+$^{58}$Ni scattering by using the Kyushu $g$-matrix folding model 
with the GHFB+AMP densities. 
\\
{\bf Results:}  
For p+$^{58}$Ni scattering, the Kyushu $g$-matrix folding model  with the GHFB+AMP densities 
reproduces  $\sigma_{\rm R}({\rm EXP})$ in $8.8 \leq E_{\rm in} \leq 81$MeV.
For $E_{\rm in}=81$MeV, we define the factor $F$ as 
$F=\sigma_{\rm R}({\rm EXP})/\sigma_{\rm R}({\rm AMP})=0.9775$.
The $F\sigma_{\rm R}({\rm AMP})$ be much the same as 
the center values of  $\sigma_{\rm R}({\rm EXP})$ in $8.8 \leq E_{\rm in} \leq 81$MeV. 
We then determine $r_{\rm m}({\rm EXP})$ from the center values of  $\sigma_{\rm R}({\rm EXP})$, 
using $\sigma_{\rm R}({\rm EXP})=C r_{m}^{2}({\rm EXP})$ with $C=r_{m}^{2}({\rm AMP})/
(F\sigma_{\rm R}({\rm AMP}))$.  
The $r_{m}({\rm EXP})$ thus obtained are averaged over $E_{\rm in}$. 
The averaged value is $r_{m}({\rm EXP})=3.697$fm. Eventually, we obtain $r_{\rm skin}({\rm EXP})=0.023$fm 
from $r_{\rm m}=3.697$fm and $r_p({\rm EXP})=3.685$fm of electron scattering. 
\end{abstract}

\maketitle


\section{Introduction and conclusion}
\label{Introduction}

{\it Background:}
A novel method for measuring nuclear reactions in inverse kinematics with stored ion beams was successfully used to extract the matter radius of $^{58}$Ni~\cite{Zamora:2017adt}. The experiment was performed at the experimental heavy-ion storage ring at the GSI facility. 
Their results determined from the differential cross section 
for $^{58}$Ni+$^{4}$He scattering are $r_m({\rm GSI})=3.70(7)$fm, $r_p({\rm GSI})=3.68$fm, 
$r_n({\rm GSI})=3.71(12)$, $r_{\rm skin}({\rm GSI})=0.03(12)$fm.

Reaction cross section $\sigma_{\rm R}$ and interaction cross sections $\sigma_{\rm I}$ are 
a standard observable to determine a central value of matter radius  $r_{\rm m}$. 
In fact,  we deduced the matter radii $r_{\rm m}$ for Ne isotopes~\cite{Sumi:2012fr} and  
for Mg isotopes~\cite{Watanabe:2014zea}. 
One can  then evaluate $r_{\rm skin}$ and $r_{\rm n}$ from the $r_{\rm m}$ and 
the $r_{\rm p}({\rm exp})$~\cite{Angeli:2013epw} of the electron scattering. 
Eventually, one can determine $r_{\rm m}$ and $r_{\rm skin}$ from the central value of  $\sigma_{\rm R}({\rm exp})$. Recently, we have determined $r_{\rm m}$ and $r_{\rm skin}$ for $^{208}$Pb~\cite{Tagami:2020bee} and 
$^{40,48}$Ca~\cite{Tagami:2020ajd,Tagami:2020pdp}, using the chiral  (Kyushu) $g$-matrix folding model with 
the densities calculated with Gogny-D1S-HFB (GHFB) with the angular momentum  projection (AMP). 
As for $^{58}$Ni, the  data on $\sigma_{\rm R}$ are available for $p$+$^{58}$Ni 
scattering~\cite{Auce:2005ks,Ingemarsson:1999sra,EliyakutRoshko:1995fn,Dicello:1967zz,Bulman:1965}

The $g$-matrix folding model is a standard way of obtaining microscopic optical potential 
for proton scattering and nucleus-nucleus scattering~\cite{Brieva-Rook,Amos,Satchler-1979,CEG, CEG07, Egashira:2014zda,Toyokawa:2014yma,Toyokawa:2015zxa,Toyokawa:2017pdd,Tagami:2019svt}. 
Applying the folding model with the Melbourne $g$-matrix~\cite{Amos} 
for  $\sigma_{\rm I}$ for Ne isotopes and $\sigma_{\rm R}$ for Mg isotopes, 
we found that $^{31}$Ne is a deformed halo nucleus~\cite{Minomo:2011bb},   
and determined the matter radii $r_{\rm m}$ for Ne isotopes~\cite{Sumi:2012fr} and  
for Mg isotopes~\cite{Watanabe:2014zea}. 

Kohno calculated the $g$ matrix  for the symmetric nuclear matter, 
using the Brueckner-Hartree-Fock method with chiral N$^{3}$LO 2NFs and NNLO 3NFs~\cite{Koh13}. 
He set $c_D=-2.5$ and $c_E=0.25$ so that  the energy per nucleon can  become minimum 
at $\rho = \rho_{0}$. 
Toyokawa {\it et al.} localized the non-local chiral  $g$ matrix into three-range Gaussian forms~\cite{Toyokawa:2017pdd}, using the localization method proposed 
by the Melbourne group~\cite{von-Geramb-1991,Amos-1994,Amos}. 
The resulting local  $g$ matrix is referred to as  ``Kyushu  $g$-matrix''.

The  Kyushu $g$-matrix folding model is successful in reproducing $\sigma_{\rm R}$ 
 and differential cross sections  $d\sigma/d\Omega$ for $^4$He scattering 
 in $E_{\rm lab}=30 \sim 200$~MeV per nucleon~\cite{Toyokawa:2017pdd}. 
The success is true for proton scattering at $E_{\rm lab}=65$~MeV~\cite{Toyokawa:2014yma}. 

{\it Proton and neutron densities used in the folding model:}
In Ref.~\cite{Tagami:2019svt}, GHFB and GHFB+AMP reproduce the one-neutron separation energy $S_{1}$ and 
the two-neutron separation energy $S_{2}$ in $^{41-58}$Ca~\cite{HP:NuDat 2.7,Tarasov2018,Michimasa2018obr}. 
We found, with  $S_{1}$ and $S_{2}$, 
that $^{64}$Ca is an even-dripline nucleus and $^{59}$Ca is an odd-dripline nucleus. 
Our results are consistent with the data~\cite{HP:NuDat 2.7} in $^{40-58}$Ca for the binding energy $E_{\rm B}$. 
This means that the proton and neutron densities are good.

{\it Aim:} 
Our aim is to to determine  matter radius $r_{m}$ and skin $r_{\rm skin}$ for $^{58}$Ni  from the 
$\sigma_{\rm R}({\rm EXP})$ of  p+$^{58}$Ni scattering by using the Kyushu $g$-matrix folding model 
with the GHFB+AMP densities. 

{\it Results:}
For p+$^{58}$Ni scattering, the Kyushu $g$-matrix folding model  with the GHFB+AMP densities 
reproduces  $\sigma_{\rm R}({\rm EXP})$ in $8.8 \leq E_{\rm in} \leq 81$MeV.
As a fine-tuning, for $E_{\rm in}=81$MeV, we define the factor $F$ as 
$F=\sigma_{\rm R}({\rm EXP})/\sigma_{\rm R}({\rm AMP})=0.9775$.
The $F\sigma_{\rm R}({\rm AMP})$ is much the same as 
the center values of  $\sigma_{\rm R}({\rm EXP})$ in $8.8 \leq E_{\rm in} \leq 81$MeV. 
We then determine $r_{\rm m}({\rm EXP})$ from the center values of  $\sigma_{\rm R}({\rm EXP})$, 
using $\sigma_{\rm R}({\rm EXP})=C r_{m}^{2}({\rm EXP})$ with $C=r_{m}^{2}({\rm AMP})/
(F\sigma_{\rm R}({\rm AMP}))$.  The $r_{m}({\rm EXP})$ thus obtained are averaged over $E_{\rm in}$. 
The averaged value is $r_{m}({\rm EXP})=3.697$fm. Eventually, we obtain $r_{\rm skin}({\rm EXP})=0.023$fm 
from $r_{\rm m}=3.697$fm and $r_p({\rm PCNP})=3.685$fm of electron scattering. 

{\it Conclusion:} 
Our conclusion is that the central value  of $r_{m}({\rm EXP})$ is 3.697fm and 
that of $r_{\rm skin}({\rm EXP)}$ is 0.023fm. 
Our results are close to with those shown in Ref.~\cite{Zamora:2017adt}.

\section{Model}
\label{Sec-Framework}

Our model is the Kyushu $g$-matrix  folding model~\cite{Toyokawa:2017pdd} 
with densities calculated with GHFB+AMP~\cite{Tagami:2019svt}.  
The folding model itself is clearly shown in Ref.~~\cite{Egashira:2014zda}. 
The Kyushu $g$-matrix is constructed from chiral interaction with the cutoff 550~MeV.

\section{Results}
\label{Results} 

\begin{figure}[H]
\begin{center}
 \includegraphics[width=0.5\textwidth,clip]{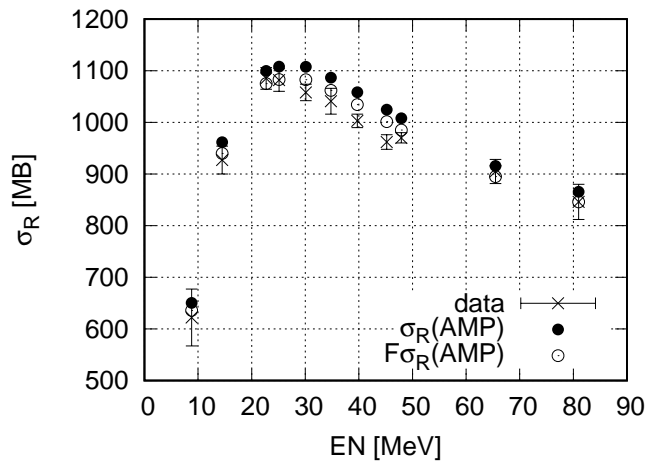}
 \caption{ 
 $E_{\rm in}$ dependence of reaction cross sections $\sigma_{\rm R}$ 
 for $p$+$^{58}$Ni scattering. 
 Closed circles denote results of the  GHFB+AMP densities, and open circles correspond 
 to $F \sigma_{\rm R}({\rm AMP})$. 
 The data (crosses) are taken from 
 Refs.~\cite{Auce:2005ks,Ingemarsson:1999sra,EliyakutRoshko:1995fn,Dicello:1967zz,Bulman:1965}.
   }
 \label{Fig-RXsec-p+Ni58}
\end{center}
\end{figure}

Figure~ \ref{Fig-RXsec-p+Ni58} shows  reaction cross sections $\sigma_{\rm R}$ 
as a function of incident energy  $E_{\rm in}$ for $p$+$^{58}$Ni scattering. 
In 2-$\sigma$ level, the Kyushu $g$-matrix folding model with the GHFB+AMP densities (closed circles) 
 reproduces  $\sigma_{\rm R}({\rm EXP})$~\cite{Auce:2005ks,Ingemarsson:1999sra,EliyakutRoshko:1995fn,Dicello:1967zz,Bulman:1965} in $8.8 \leq E_{\rm in} \leq 81$MeV; note that the data has high accuracy of 2.7~\%. 

Now, we introduce the fine-tuning factor $F$. 
We consider $E_{\rm in}=81$MeV, since total cross section of nucleon-nucleon scattering is smallest 
in $8.8 \leq E_{\rm in} \leq 81$MeV. 
For  $E_{\rm in}=81$MeV, we define the factor $F$ as 
$F=\sigma_{\rm R}({\rm EXP})/\sigma_{\rm R}({\rm AMP})=0.9775$.
The $F\sigma_{\rm R}({\rm AMP})$ (open circles) are much the same as 
the center values of  $\sigma_{\rm R}({\rm EXP})$ in $8.8 \leq E_{\rm in} \leq 81$MeV. 
We then determine $r_{\rm m}({\rm EXP})$ from the center values of  $\sigma_{\rm R}({\rm EXP})$, 
using $\sigma_{\rm R}({\rm EXP})=C r_{m}^{2}({\rm EXP})$ with $C=r_{m}^{2}({\rm AMP})/
(F\sigma_{\rm R}({\rm AMP}))$.  
The $r_{m}({\rm EXP})$ thus obtained are averaged over $E_{\rm in}$. 
The averaged value is $r_{m}({\rm EXP})=3.697$fm. We then obtain $r_{\rm skin}({\rm EXP})=0.023$fm 
and $r_{n}({\rm EXP})=3.708$~fm from 
$r_{\rm m}({\rm EXP})=3.697$~fm and $r_p({\rm EXP})=3.685$fm of the electron scattering. 
Our results agree with $r_m({\rm GSI})=3.70(7)$fm, $r_p({\rm GSI})=3.68$fm, 
$r_n({\rm GSI})=3.71(12)$, $r_{\rm skin}({\rm GSI})=0.03(12)$fm.

\noindent
\appendix

\noindent
\begin{acknowledgments}
We would like to thank Dr. Toyokawa for providing his code. 
\end{acknowledgments}



\bibliographystyle{prsty}

\end{document}